\pdfoutput=1
\documentclass[
 reprint,
 aps,
 prl,
 superscriptaddress,
 nofootinbib,
 floatfix,
 amsmath,amssymb,
 preprintnumbers
]{revtex4-2}

\usepackage[colorlinks=true,linkcolor=blue,citecolor=blue,urlcolor=blue]{hyperref}
\usepackage[T1]{fontenc}
\usepackage{graphicx}
\usepackage{bm,booktabs}
\usepackage{amssymb}
\usepackage{dsfont}
\usepackage{placeins}
\usepackage{comment}
\usepackage{orcidlink}
\usepackage{tikz}
\usetikzlibrary{arrows.meta,decorations.markings}
\newcommand{\mr}[1]{\mathrm{#1}}

\begin{document}
\preprint{OU-HET-1326}

\title{A non-invertible Fibonacci selection rule for two-Higgs-doublet models}

\author{Yu Hamada\,\orcidlink{0000-0002-0227-5919}}
\email{yu.hamada@het.phys.sci.osaka-u.ac.jp}
\affiliation{
Department of Physics, The University of Osaka,
Machikaneyama-cho, Toyonaka, Osaka 560-0043, Japan
}

\begin{abstract}
We study two-Higgs-doublet models, extensions of the Standard Model (SM) containing two Higgs doublets rather than one. 
Tree-level Higgs-mediated interactions that change quark or lepton flavor are conventionally suppressed by a discrete $\mathbb{Z}_2$ symmetry. 
Motivated by string compactifications, we instead impose a non-invertible Fibonacci selection rule at a matching scale $\Lambda$ for that purpose. 
This rule also forbids the independent mass parameter that normally allows the additional Higgs particles to become arbitrarily heavy, while renormalization-group evolution regenerates it below $\Lambda$.
An allowed $\mathbb{Z}_2$-breaking quartic interaction simultaneously avoids stable cosmological domain walls.
In the exact alignment limit, requiring perturbative unitarity and boundedness from below up to the matching scale
bounds the masses of the three non-SM Higgs states; for $\Lambda=10^{19}\,\mathrm{GeV}$, they lie below about $150\,\mathrm{GeV}$. 
The scanned parameter space predicts a suppressed Higgs diphoton signal strength within the projected reach of the HL-LHC. 
This illustrates how a stringy non-invertible selection rule can leave testable signatures in Higgs physics.
\end{abstract}

\maketitle

\textit{Introduction.}---Generalized symmetries extend the conventional notion of symmetry beyond ordinary group actions~\cite{Gaiotto:2014kfa}. 
In particular, non-invertible symmetries obey fusion rules that need not form a group~\cite{Thorngren:2019iar,Koide:2021zxj,Choi:2021kmx,Kaidi:2021xfk,Cordova:2022ieu} and can give rise to selection rules unavailable from conventional group charges~\cite{Kaidi:2024wio,Suzuki:2025oov,Kobayashi:2024cvp,Suzuki:2025bxg}.
Such selection rules can arise naturally from string compactifications~\cite{Heckman:2024obe,Kaidi:2024wio,Kobayashi:2024yqq,Dong:2025jra,Kobayashi:2025ocp,Dong:2026iwa}, raising the possibility that they leave distinctive low-energy signatures.

As a concrete example, we consider the Fibonacci fusion algebra in this Letter:
\begin{align}
 \mathbf{1}\times \mathbf{1}&=\mathbf{1},
&
 \mathbf{1}\times \tau&=\tau,
\nonumber\\
 \tau\times \tau&=\mathbf{1}\oplus \tau.
 \label{eq:fib-rule}
\end{align}
Here $\mathbf{1}$ is the identity object and $\tau$ is self-dual but has no inverse.
An operator is allowed by the associated Fibonacci selection rule (FSR) when the identity object appears in the fusion product of the labels carried by its fields.
A crucial difference from an ordinary $\mathbb{Z}_2$ is that a single $\tau$ field is forbidden, while three $\tau$ fields can form an invariant because $\tau^3$ contains the identity.

In this Letter, we show that this distinction has a direct consequence in the two-Higgs-doublet model (2HDM): it realizes natural flavor conservation while imposing a qualitatively different structure on the scalar potential, leading to predictive relations for the additional particles.

\textit{Fibonacci two-Higgs-doublet model.}---The 2HDM is a simple and widely studied extension of the SM containing a second Higgs doublet~\cite{Branco:2011iw}.
Conventionally, tree-level flavor-changing neutral currents (FCNCs) are avoided by imposing a softly broken $\mathbb{Z}_2$ symmetry under which each right-handed fermion couples to only one Higgs doublet~\cite{Glashow:1976nt,Paschos:1976ay,Branco:2011iw}. Although this provides an efficient phenomenological solution, the origin of the $\mathbb{Z}_2$ symmetry is left unexplained within the low-energy theory. Here we instead explore a top-down alternative in which the Yukawa structure is inherited from UV dynamics.

The FSR provides a suitable framework for this purpose. As noted above, it can arise from UV constructions such as string compactifications and can realize natural flavor conservation without imposing the conventional $\mathbb{Z}_2$ symmetry. We call the resulting theory the “Fibonacci two-Higgs-doublet model” (F2HDM).

\begin{table}[tbp]
\centering
\begin{tabular}{ccccc}
\toprule
Type & $u_R$ & $d_R$ & $e_R$ & Allowed doublets \\
\midrule
I & $\tau$ & $\tau$ & $\tau$
  & all fermions couple to $\Phi_2$ \\
II & $\tau$ & $\mathbf{1}$ & $\mathbf{1}$
  & $u$ to $\Phi_2$, $d,e$ to $\Phi_1$ \\
X & $\tau$ & $\tau$ & $\mathbf{1}$
  & quarks to $\Phi_2$, leptons to $\Phi_1$ \\
Y & $\tau$ & $\mathbf{1}$ & $\tau$
  & $u,e$ to $\Phi_2$, $d$ to $\Phi_1$ \\
\bottomrule
\end{tabular}
\caption{
Examples of Fibonacci labels that reproduce the usual natural-flavor-conserving
2HDM Yukawa structures.  The left-handed doublets are assigned to
$\mathbf{1}$.
}
\label{tab:yukawa}
\end{table}

We introduce two Higgs doublets $\Phi_1$ and $\Phi_2$ labeled by
\begin{equation}
  \Phi_1\sim \mathbf{1},\qquad \Phi_2\sim \tau,
  \label{eq:2hdm-scalar-assignment}
\end{equation}
and take the left-handed quark and lepton doublets to carry the label $\mathbf{1}$.
For a right-handed fermion $f_R$, the two possible labels allow the following Yukawa structures:
\begin{align}
f_R\sim \mathbf{1} \, &\Rightarrow
 \bar{F}_L \Phi_1 f_R \quad {\rm allowed}, \\
f_R\sim \tau \, &\Rightarrow
 \bar{F}_L \Phi_2 f_R \quad {\rm allowed},
\end{align}
with $F_L$ denoting a left-handed quark or lepton doublet
and $\Phi_i$ replaced by $\tilde{\Phi}_i$ for up-type quarks.  

Therefore, at tree level, each right-handed fermion species couples to only one Higgs doublet, realizing natural flavor conservation and eliminating tree-level FCNCs by the same logic as in the usual $\mathbb{Z}_2$ models.
The familiar four Yukawa types of the 2HDM can be reproduced by the assignments in Table~\ref{tab:yukawa}.
(The other possible Yukawa structures will be discussed later.)

By contrast, the FSR has distinctive consequences for the scalar sector.
To make the effect explicit, we begin with the
most general 2HDM scalar potential,
\begin{align}
V=& \,
 m_{11}^2\,\Phi_1^\dagger\Phi_1
 +m_{22}^2\,\Phi_2^\dagger\Phi_2
 -\left[m_{12}^2\,\Phi_1^\dagger\Phi_2+{\rm h.c.}\right]
\nonumber\\
&+\frac{\lambda_1}{2}(\Phi_1^\dagger\Phi_1)^2
 +\frac{\lambda_2}{2}(\Phi_2^\dagger\Phi_2)^2
\nonumber\\
&+\lambda_3(\Phi_1^\dagger\Phi_1)(\Phi_2^\dagger\Phi_2)
 +\lambda_4(\Phi_1^\dagger\Phi_2)(\Phi_2^\dagger\Phi_1)
\nonumber\\
&+\Bigg[\frac{\lambda_5}{2}(\Phi_1^\dagger\Phi_2)^2 + \lambda_6(\Phi_1^\dagger\Phi_1)(\Phi_1^\dagger\Phi_2) \nonumber \\
&  \hspace{3em}+ \lambda_7(\Phi_2^\dagger\Phi_2)
   (\Phi_1^\dagger\Phi_2)+{\rm h.c.}\Bigg],
 \label{eq:2hdm-potential}
\end{align}
and restrict our analysis to the CP-conserving case by taking all parameters to
be real.
We assume a neutral, CP-conserving vacuum with real nonzero $v_1$ and $v_2$,
$\langle\Phi_1\rangle=(0,v_1/\sqrt{2})^T$ and
$\langle\Phi_2\rangle=(0,v_2/\sqrt{2})^T$,
where $v^2\equiv v_1^2+v_2^2\simeq(246\,\mathrm{GeV})^2$
and $t_\beta\equiv\tan\beta=v_2/v_1$.

With the labeling in Eq.~\eqref{eq:2hdm-scalar-assignment}, the FSR forbids
\begin{equation}
  m_{12}^2 \Phi_1^\dagger \Phi_2 +{\rm h.c.},\qquad
  \lambda_6 (\Phi_1^\dagger\Phi_1)(\Phi_1^\dagger\Phi_2)+{\rm h.c.},
\end{equation}
while allowing
\begin{equation}
  \lambda_7(\Phi_2^\dagger\Phi_2)(\Phi_1^\dagger\Phi_2)+{\rm h.c.}
\end{equation}
in the potential~\eqref{eq:2hdm-potential}.
This pattern cannot arise from the exact group-like $\mathbb{Z}_2$: $\Phi_1\to \Phi_1$, $\Phi_2\to -\Phi_2$.

The FSR is a tree-level selection rule and is not preserved by radiative corrections.
In other words, it is realized at a UV energy scale $\Lambda$ (dubbed the matching scale),
typically the string compactification scale~\cite{Kobayashi:2024yqq,Dong:2025jra,Kobayashi:2025ocp,Dong:2026iwa},
and renormalization-group (RG) evolution then induces the above forbidden operators at energy scales below $\Lambda$,
a phenomenon known as ``groupification''~\cite{Kaidi:2024wio}.
Since the Fibonacci fusion algebra is groupified into the trivial group $1$,
all the operators can in principle be generated radiatively.

Accordingly, we implement the selection rule as
boundary conditions on the parameters in
Eq.~\eqref{eq:2hdm-potential} at the matching scale $\Lambda$, rather than treating it as an exact symmetry at all scales.
Equivalently, the boundary conditions are
$m_{12}^2(\Lambda)=\lambda_6(\Lambda)=0$, whereas $\lambda_7(\Lambda)$ need not vanish.
Even though they are induced at the EW scale $v$ through radiative corrections,
the F2HDM retains considerable predictive power as seen below.
For a related 2HDM scenario where a non-invertible symmetry controls $m_{12}^2$, which is generated by UV instantons,
see Ref.~\cite{Delgado:2024pcv}.

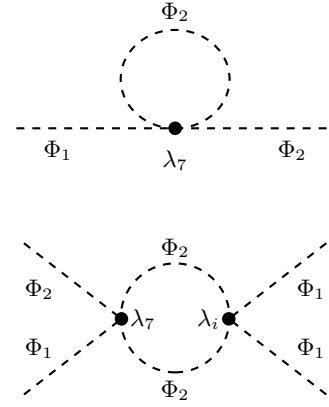
\begin{figure}[tbp]
\centering
\begin{tikzpicture}[
  x=1cm,y=1cm,
  scalar/.style={dashed,line width=0.8pt},
  vertex/.style={circle,fill=black,inner sep=1.8pt}
]
  \coordinate (v) at (0,0);
  \draw[scalar] (-2.1,0) -- (v);
  \draw[scalar] (v) -- (2.1,0);
  \draw[scalar] (v)
    arc[start angle=-90,end angle=270,x radius=0.72cm,y radius=0.65cm];
  \node[vertex] at (v) {};
  \node[below] at (-1.55,-0.05) {$\Phi_1$};
  \node[below] at (1.55,-0.05) {$\Phi_2$};
  \node[above] at (0,1.30) {$\Phi_2$};
  \node[below] at (0,-0.18) {$\lambda_7$};
\end{tikzpicture} \\[2em]
\begin{tikzpicture}[
  x=1cm,y=1cm,
  scalar/.style={dashed,line width=0.8pt},
  vertex/.style={circle,fill=black,inner sep=1.8pt}
]
  \coordinate (v1) at (-0.71,0);
  \coordinate (v2) at (0.71,0);
  \draw[scalar] (-2.,1) -- (v1);
  \draw[scalar] (-2.,-1) -- (v1);
  \draw[scalar] (2.,1) -- (v2);
  \draw[scalar] (2.,-1) -- (v2);
  \draw[scalar] (0,-0.707)
    arc[start angle=-90,end angle=270,x radius=0.72cm,y radius=0.72cm];
  \node[vertex] at (v1) {};
  \node[vertex] at (v2) {};
  \node[below] at (-1.8,-0.15) {$\Phi_1$};
  \node[above] at (-1.8,0.15) {$\Phi_2$};
  \node[below] at (1.8,-0.15) {$\Phi_1$};
  \node[above] at (1.8,0.15) {$\Phi_1$};
  \node[above] at (0,0.7) {$\Phi_2$};
  \node[below] at (0,-0.7) {$\Phi_2$};
  \node[left] at (v2) {$\lambda_i$};
  \node[right] at (v1) {$\lambda_7$};
\end{tikzpicture}
\caption{One-loop Feynman diagrams generated by the
$\lambda_7(\Phi_2^\dagger\Phi_2)(\Phi_1^\dagger\Phi_2)$ vertex.
The coupling $\lambda_i$ denotes $\lambda_3,\lambda_4,$ or $\lambda_5$. }
\label{fig:radiative-m12-l6}
\end{figure}

The FSR at the scale $\Lambda$ has an immediate consequence.  
The standard decoupling mass parameter~\cite{Gunion:2002zf}
\begin{equation}
  M^2=\frac{m_{12}^2}{\sin\beta\cos\beta}\label{eq:M}
\end{equation}
vanishes: $M^2(\Lambda)=0$
without causing the domain-wall problem associated with an exact, spontaneously broken $\mathbb{Z}_2$ symmetry~\cite{Eto:2018hhg,Battye:2020jeu}
because a generic nonzero $\lambda_7(\Lambda)$ explicitly breaks the $\mathbb{Z}_2$ symmetry.
At lower scales, however, $m_{12}^2$ and $\lambda_6$ are radiatively generated by non-zero $\lambda_7$ as shown in Fig.~\ref{fig:radiative-m12-l6}.
Once generated, they also contribute to their subsequent RG evolution.
Nevertheless, the FSR still has a significant impact on the mass spectrum at the EW scale.

\begin{figure*}[htbp]
\centering
\includegraphics[width=0.32\textwidth]{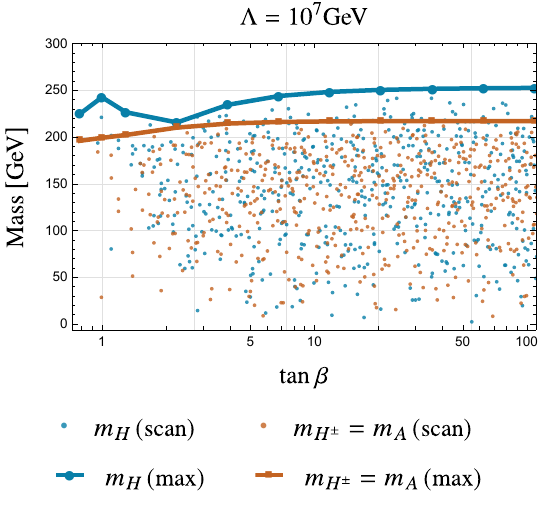} 
\includegraphics[width=0.32\textwidth]{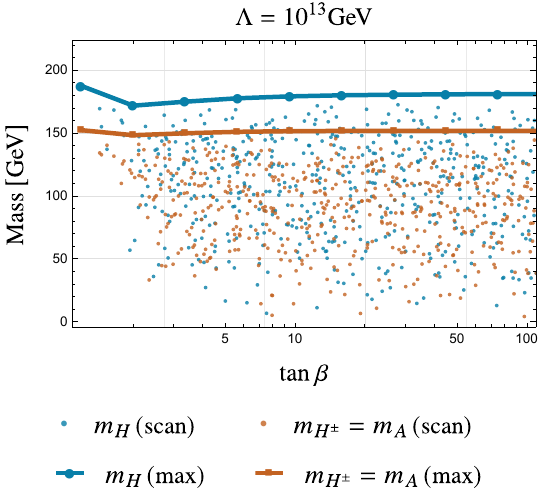} 
\includegraphics[width=0.32\textwidth]{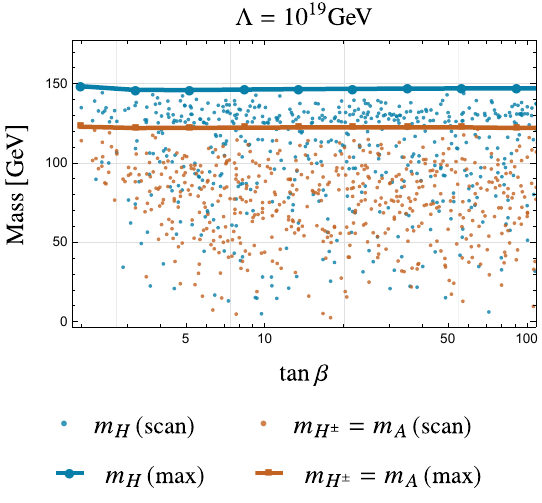}
\caption{
Plots for masses of the non-SM Higgs bosons $m_H$ (blue) and $m_{H^\pm}=m_A$ (orange) versus $\tan\beta$ in the F2HDM.
The scatter plots show randomly generated parameter points 
that solve the RG equations with the correct boundary conditions and satisfy the theoretical constraints (PU + BFB)
for $\Lambda=10^{7}\, \mr{GeV}$ (left), $10^{13}\, \mr{GeV}$ (middle), and $10^{19}\, \mr{GeV}$  (right).  
The points with solid lines show their maximum values consistent with the RGE and the theoretical constraints.
}
\label{fig:2hdm-pu-nondegenerate}
\end{figure*}

To see this, we now consider the physical Higgs masses.
The physical scalar spectrum consists of
the charged Higgs bosons $H^\pm$, the neutral CP-odd Higgs boson $A$, and the two neutral CP-even Higgs bosons $h$ and $H$.
In the Higgs alignment limit~\cite{Branco:2011iw} used below,
in which $h$ is the SM-like state,\footnote{In our convention, we call the SM-like Higgs boson $h$ and the other neutral CP-even state $H$ in the alignment limit even when $H$ is lighter.}
the masses of the other states can be written as
\begin{align}
 m_A^2
 &=M^2-v^2
 \left(
   \lambda_5+\frac{\lambda_6}{2}t_\beta^{-1}
             +\frac{\lambda_7}{2}t_\beta
 \right),
 \label{eq:mA}
\\
 m_{H^\pm}^2
 &=M^2-\frac{v^2}{2}
 \left(
   \lambda_4+\lambda_5+\lambda_6t_\beta^{-1}+\lambda_7t_\beta
 \right),
 \label{eq:mHp}
\\
 m_H^2
 &=M^2+\frac{v^2}{t_\beta^2}
 \left(
   \lambda_1-\frac{m_h^2}{v^2}
   +\frac{3}{2}\lambda_6t_\beta
   -\frac{1}{2}\lambda_7t_\beta^3
 \right).
 \label{eq:mH-alignment}
\end{align}
Therefore, all non-SM Higgs masses are controlled by
quartic interactions $\lambda_i$, $\tan\beta$, and 
$M^2$ defined by Eq.~\eqref{eq:M}.
As a consequence of the FSR, $M^2$ is no longer an arbitrary parameter but is determined by RG evolution,
which implies that we can obtain a non-trivial prediction on the non-SM Higgs masses.

\textit{Renormalization-group evolution.}---To obtain quantitative predictions, we solve the one-loop RG equations in the $\overline{\mr{MS}}$ scheme for the running parameters,
\begin{align}
 \frac{d}{d\log\mu}m_{ij}^2(\mu) &= \beta_{m_{ij}^2}(m^2,\lambda,g,Y_t),\label{eq:RGeq1} \\
 \frac{d}{d\log\mu}\lambda_{k}(\mu) &= \beta_{\lambda_{k}}(m^2,\lambda,g,Y_t),\label{eq:RGeq2}
\end{align}
with $i,j\in\{1,2\}$ and $k\in\{1,\ldots,7\}$,
where $m^2$, $\lambda$, and $g$ collectively denote the quadratic parameters,
quartic couplings, and SM gauge couplings, respectively, and $Y_t$ denotes the top-quark Yukawa coupling with $\Phi_2$.
The one-loop beta functions $\beta_{m_{ij}^2}$ and $\beta_{\lambda_k}$ are given in Appendix~\ref{sec:rge}.
Together with Eqs.~\eqref{eq:RGeq1} and \eqref{eq:RGeq2}, we simultaneously evolve $g$ and $Y_t$ at one loop, while neglecting the other fermion Yukawa couplings.
We impose the boundary conditions $m_{12}^2(\Lambda)=\lambda_6(\Lambda)=0$ at the matching scale $\mu=\Lambda$. 
At the EW scale $\mu=v$, we impose $m_h=125\,\mathrm{GeV}$, $v=246\,\mathrm{GeV}$, exact alignment, and the degeneracy condition $m_{H^\pm}=m_A$, the latter suppressing contributions to electroweak precision observables.
These conditions leave four independent parameters, including $\tan\beta$.

In addition, we impose two theoretical constraints on the running quartic couplings all the way from $\mu=v$ to $\mu=\Lambda$:
perturbative unitarity (PU)~\cite{Lee:1977eg} and boundedness from below (BFB) of the scalar potential.
See Appendix~\ref{sec:constraint} for the explicit form of these conditions in the F2HDM.
Because the seven quartic couplings are coupled through the RG equations, 
at least some of them other than $\lambda_6$ typically grow toward the UV, 
so that the theoretical constraints tend to become most restrictive near the matching scale $\mu=\Lambda$. 
Requiring PU and BFB up to $\Lambda$ therefore restricts the quartic couplings already at the EW scale, leading to a light non-SM Higgs spectrum in Eqs.~\eqref{eq:mA}-\eqref{eq:mH-alignment}.
Related RG analyses have been used to derive Higgs-mass bounds in 2HDMs~\cite{Kanemura:1999xf} and to study the low-energy consequences of high-scale symmetry conditions~\cite{Aiko:2020atr}.

\textit{Higgs-mass bounds.}---We first perform random scans for the model parameters with three different values of the matching scale, $\Lambda=10^7,10^{13}$, and $10^{19}\,\mathrm{GeV}$. 
We generate EW-scale parameter points whose RG evolution satisfies the boundary conditions at both $\mu=v$ and $\mu=\Lambda$. 
We then retain the points that satisfy PU and BFB throughout the RG evolution and evaluate the corresponding values of $\tan\beta$, $m_H$, and $m_{H^\pm}=m_A$.
The allowed points are shown as scatter plots in Fig.~\ref{fig:2hdm-pu-nondegenerate} (blue for $m_H$ and orange for $m_{H^\pm}=m_A$), and each parameter point therefore contributes one blue point for $m_H$ and one orange point for $m_{H^\pm}=m_A$.
The plots exhibit clear upper envelopes. 
To determine these upper bounds more precisely, we independently maximize $m_H$ and $m_{H^\pm}=m_A$ at fixed $\tan\beta$, subject to the RG boundary conditions and the PU and BFB constraints along the RG flow.
The resulting maxima are shown as the connected solid curves in Fig.~\ref{fig:2hdm-pu-nondegenerate}  and are consistent with the envelopes of the random-scan points.

The large-$\tan\beta$ region in Fig.~\ref{fig:2hdm-pu-nondegenerate} highlights a further
distinction from the exact-$\mathbb{Z}_2$ 2HDM.
In the latter, $m_{12}^2=\lambda_6=\lambda_7=0$, and
Eq.~\eqref{eq:mH-alignment} gives
$\lambda_1=(m_h^2+m_H^2\tan^2\beta)/v^2$.
PU therefore bounds $\tan\beta$ at fixed nonzero $m_H$.
In the F2HDM, the allowed $\lambda_7$ interaction relaxes
this mass--$\tan\beta$ correlation, permitting large
$\tan\beta$ with light non-SM Higgs masses.
Thus, the same quartic interaction both avoids the
$\mathbb{Z}_2$ domain-wall problem and opens the large-$\tan\beta$
regime, without introducing the independently adjustable
decoupling mass parameter.

The resulting mass bounds are striking: 
the non-SM Higgs masses should be quite light.
For instance, they should be lighter than about $150\,\mr{GeV}$ for $\Lambda=10^{19}\, \mr{GeV}$.
The flavor analyses in
Refs.~\cite{Misiak:2020vlo,Li:2024kpd} yield a lower bound of
$m_{H^\pm}\gtrsim800\,\mathrm{GeV}$ at 95\% C.L. in type-II and type-Y (F)2HDMs,
excluding the light-Higgs spectra considered here in the
type-II and type-Y.
In contrast, these flavor constraints do not generically exclude
such light additional Higgs bosons in the type-I and type-X
F2HDMs~\cite{Li:2024kpd}.
Nevertheless, the type-X F2HDM is already largely excluded in the light-Higgs mass region considered here by existing collider searches,
most notably the CMS search for $Z^*\to HA\to4\tau$~\cite{CMS:2026ntm}.

\begin{figure}[tbp]
\centering

\begin{tikzpicture}[
  scale=0.85,
  x=1cm, y=1cm,
  font=\small,
  line cap=round,
  line join=round,
  scalar/.style={
    line width=0.8pt,
    dash pattern=on 3pt off 2pt
  },
  fermion/.style={
    line width=0.85pt,
    postaction={decorate},
    decoration={markings,
      mark=at position 0.52 with {
        \arrow{Latex[length=1.65mm,width=1.15mm]}
      }
    }
  },
  vertex/.style={
    circle,
    fill=black,
    inner sep=1.45pt
  },
  lab/.style={inner sep=1.5pt}
]

  \def\R{1.65}
  \begin{scope}[shift={(-1.85,0)}]
    \coordinate (yu1)  at ( 90:\R);
    \coordinate (yuc1) at ( 45:\R);
    \coordinate (yu2)  at (  0:\R);
    \coordinate (yuc2) at (-45:\R);
    \coordinate (yd)   at (-90:\R);
  \end{scope}

  \coordinate (l7)   at (1.85,0);
  \coordinate (qin)  at (-3.00, \R);
  \coordinate (din)  at (-3.00,-\R);
  \coordinate (hout) at (3.05,0);

  \draw[fermion] (din) -- (yd);

  \draw[fermion] (yd)
    arc[start angle=-90,end angle=-45,radius=\R cm];
  \draw[fermion] (yuc2)
    arc[start angle=-45,end angle=0,radius=\R cm];
  \draw[fermion] (yu2)
    arc[start angle=0,end angle=45,radius=\R cm];
  \draw[fermion] (yuc1)
    arc[start angle=45,end angle=90,radius=\R cm];

  \draw[fermion] (yu1) -- (qin);

  \draw[scalar] (yu1)
    -- node[lab,left=4pt] {$\Phi_2$} (yd);

  \draw[scalar] (yuc1)
    -- node[lab,pos=0.53,above=3pt] {$\Phi_2$} (l7);

  \draw[scalar] (yu2)
    -- node[lab,pos=0.28,above=1pt] {$\Phi_2^\dagger$} (l7);

  \draw[scalar] (yuc2)
    -- node[lab,pos=0.53,below=3pt] {$\Phi_2$} (l7);

  \draw[scalar] (l7)
    -- node[lab,pos=0.58,above=3pt] {$\Phi_1$} (hout);

  \foreach \p in {yu1,yuc1,yu2,yuc2,yd,l7}{
    \node[vertex] at (\p) {};
  }

  \node[lab,left=4pt] at (qin) {$\bar Q_L$};
  \node[lab,left=4pt] at (din) {$d_R$};

  \node[lab,above=5pt] at (yu1)  {$Y_u$};
  \node[lab,left=3pt]  at (yuc1) {$Y_u^\dagger$};
  \node[lab,left=3pt]  at (yu2)  {$Y_u$};
  \node[lab,left=3pt]  at (yuc2) {$Y_u^\dagger$};
  \node[lab,below=5pt] at (yd)   {$Y_d$};
  \node[lab,below=6pt] at (l7)   {$\lambda_7$};
\end{tikzpicture}
\caption{Representative three-loop diagram inducing the Yukawa coupling $\bar Q_L\Phi_1d_R$,
which is absent at tree level in the type-I and type-X F2HDMs.}
\label{fig:radiative-yukawa}
\end{figure}
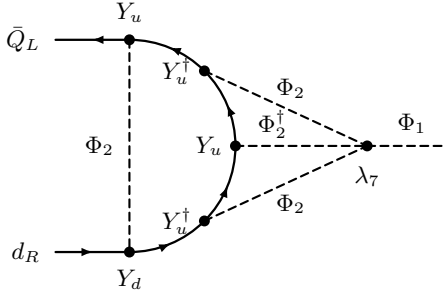

\textit{Higgs-mediated FCNC.}---Because the FSR is imposed as a boundary condition rather
than as an exact symmetry at all scales, 
the Yukawa structure also receives radiative corrections.
For definiteness, we consider the type-I and type-X F2HDMs.
In these cases, RG evolution induces the coupling of down-type quarks to $\Phi_1$. 
The leading RG-induced contribution to the flavor-off-diagonal Yukawa couplings arises at three loops
as illustrated in Fig.~\ref{fig:radiative-yukawa} and discussed in Ref.~\cite{Herren:2017uxn}. 
In the down-quark mass basis, the off-diagonal entries of this Yukawa matrix mediate FCNCs. 
For example, the $(2,3)$ entry, relevant to $b\leftrightarrow s$ transitions, is approximately
\begin{align}
 (Y^d_1)_{23} & \simeq \frac{1}{(16\pi^2)^3} \frac{4\sqrt{2}\lambda_7}{\sin^5\beta} V_{ts}^\ast \, V_{tb}\, \frac{m_t^4 m_b}{v^5} \, \log \frac{\Lambda}{v},
\end{align}
with $V_{ts}$ and $V_{tb}$ elements of the Cabibbo--Kobayashi--Maskawa matrix, and $m_t$ and $m_b$ the top-quark and bottom-quark masses, respectively.
We have ignored $\mathcal{O}(1)$ factors.
To illustrate its size, 
we find
$|(Y^d_1)_{23}|\sqrt{v^2/(m_b m_s)}= \mathcal{O}\left(10^{-6}\right)$ for
$\lambda_7\sim\sin\beta\sim1$ and $\log(\Lambda/v)\sim 10$.
Such flavor-changing couplings can contribute to observables such as $B_s-\bar B_s$ mixing~\cite{Crivellin:2013wna}.

\textit{Higgs signal strength.}---The charged Higgs boson $H^{\pm }$ modifies the loop-induced decay of the SM-like Higgs boson into two photons ($h \to \gamma\gamma$). 
The signal strength of this process $\mu_{\gamma\gamma}$,
defined as
\begin{align}
 \mu_{\gamma\gamma} \equiv \frac{\sigma(pp\rightarrow h) \mr{BR}(h \rightarrow \gamma \gamma)}{\sigma(pp \rightarrow h)_\mr{SM}\mr{BR}(h \rightarrow \gamma \gamma)_\mr{SM}},
\end{align}
depends on $M^2(v)$, $m_{H^\pm}$, and $m_{H}$ through the branching ratio $\mr{BR}(h \rightarrow \gamma \gamma)$.
We here assume that the production cross section $\sigma(pp\rightarrow h)$ is the same as that of the SM because of the alignment limit.
ATLAS and CMS measure $\mu_{\gamma\gamma}$ with high precision~\cite{ATLAS:2025qxq,CMS:2026zxu}, placing strict bounds on the model parameter space.

\begin{figure}[tbp]
\centering
\includegraphics[width=0.45\textwidth]{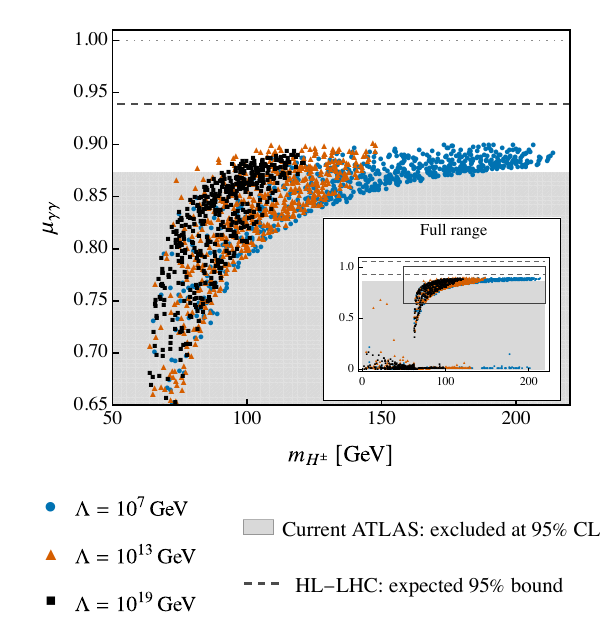} 
\caption{
Scatter plots for masses of the charged Higgs boson $m_{H^\pm}$ versus the signal strength for $h \to \gamma\gamma$ ($\mu_{\gamma\gamma}$) in the F2HDM,
corresponding to the points shown in Fig.~\ref{fig:2hdm-pu-nondegenerate}.
The gray region is excluded with $95\%\,\mr{C.L.}$ by the measurement of $\mu_{\gamma\gamma}$ by ATLAS~\cite{ATLAS:2025qxq}
 while the dashed lines represent boundaries of the expected exclusion by HL-LHC~\cite{CMS:2018qgz}.
The dotted line indicates the SM value $\mu_{\gamma\gamma}=1$.
}
\label{fig:mu_h_2gam}
\end{figure}

Figure~\ref{fig:mu_h_2gam} shows the scan points from Fig.~\ref{fig:2hdm-pu-nondegenerate} for three values of $\Lambda$ in the $(m_{H^\pm},\mu_{\gamma\gamma})$ plane.
The gray shaded region is excluded at 95\% C.L. by the current ATLAS measurement ($\mu_{\gamma\gamma}=1.05\pm0.09$)~\cite{ATLAS:2025qxq},
while the dashed line marks the projected 95\% C.L. exclusion boundary at the HL-LHC ($\mu_{\gamma\gamma}=1.0\pm0.031$)~\cite{CMS:2018qgz}.
We construct the corresponding 95\% confidence intervals from the quoted $1\sigma$ uncertainties using a Gaussian approximation.

Note that this analysis applies to all of the Yukawa types shown in Table~\ref{tab:yukawa}
in the approximation of retaining only the top-quark Yukawa coupling in the RG evolution.
Some scan points survive the adopted current constraint, whereas all displayed points lie inside the projected HL-LHC exclusion (i.e., below the lower dashed line),
so that the F2HDM can be tested by HL-LHC.

\textit{Discussion.}---Several directions remain to explore the generality of these results.
There are four additional Yukawa structures in the F2HDM beyond those listed in Table~\ref{tab:yukawa}, obtained by flipping the labels of all right-handed fermions, $\mathbf{1}\leftrightarrow\tau$. 
Their RG evolution differs because the top quark then couples to the other Higgs doublet. We expect the qualitative mechanism discussed above to remain similar, while a dedicated analysis is left for future work.

Although we have assumed exact alignment in this Letter, understanding its origin within the non-invertible framework remains an interesting direction.
In addition, the CP-invariant scalar potential assumed here may arise from a generalized CP symmetry associated with non-invertible selection rules~\cite{Kobayashi:2025wty}.

Our results demonstrate how non-invertible selection rules imposed in the UV can yield quantitative and testable predictions for extended Higgs sectors through RG evolution, opening a direction for analogous studies in other scalar sectors.

\textit{Acknowledgments.}---
Y.H. thanks Hajime Otsuka for inspiring discussions during the workshop ``Progress in Particle Physics 2026'' (YITP-W-26-9) at the Yukawa Institute for Theoretical Physics, Kyoto University.
Y.H. also thanks Naoto Kan, Shinya Kanemura and Ryosuke Sato for useful discussions.
The work of Y.H. is supported by the Japan Society for the Promotion of Science
(JSPS) KAKENHI Grant No. JP26K17155.

\bibliography{references}
\newpage 
\onecolumngrid
\begin{center}
{\large\bfseries End Matter}
\end{center}
\appendix
\setcounter{secnumdepth}{1}

\refstepcounter{section}
\label{sec:constraint}
\noindent\textit{Appendix~\thesection:Theoretical constraints.}---The two-body scalar scattering amplitude can be expanded in partial waves as
$a^\ell_{fi}$
\begin{align}
 {\cal M}_{fi}
 &=
 16\pi\sum_\ell (2\ell+1)P_\ell(\cos\theta)\,
 a^\ell_{fi} \, ,
\end{align}
where $P_\ell$ is the Legendre polynomial,
and hence perturbative unitarity (PU) can be imposed in $a^\ell_{fi}$:
\begin{align}
   |\mathrm{Re}\,a_\ell^{(i)}| &\leq \frac{1}{2}, \qquad a_\ell^{(i)} \in \mathrm{eig}\, a^\ell_{fi} \, .
 \label{eq:pu-convention}
\end{align}
In the high-energy limit considered below, only the $\ell=0$ amplitudes from the quartic interactions are relevant.

With this convention, 
the full PU bound in the 2HDM may be written as~\cite{Kanemura:1993hm,Akeroyd:2000wc,Ginzburg:2005dt,Kanemura:2015ska}
\begin{equation}
 \mathcal{U}\equiv \max\left\{
 |\mathrm{eig}\,X|,\,
 |\mathrm{eig}\,Y|,\,
 |\mathrm{eig}\,Z|,\,
 |\lambda_3-\lambda_4|
 \right\}\leq 8\pi ,
 \label{eq:2hdm-pu-blocks}
\end{equation}
\begingroup
\setlength{\arraycolsep}{2pt}
\begin{align}
X={}&
\scalebox{0.90}{$
\begin{pmatrix}
3\lambda_1 & 2\lambda_3+\lambda_4 & 3\sqrt{2}\lambda_6 & 0\\
2\lambda_3+\lambda_4 & 3\lambda_2 & 3\sqrt{2}\lambda_7 & 0\\
3\sqrt{2}\lambda_6 & 3\sqrt{2}\lambda_7 & \lambda_3+2\lambda_4+3\lambda_5 & 0\\
0 & 0 & 0 & \lambda_3+2\lambda_4-3\lambda_5
\end{pmatrix}
$},
\nonumber\\
Y={}&
\begin{pmatrix}
\lambda_1 & \lambda_4 & \sqrt{2}\lambda_6 & 0\\
\lambda_4 & \lambda_2 & \sqrt{2}\lambda_7 & 0\\
\sqrt{2}\lambda_6 & \sqrt{2}\lambda_7 & \lambda_3+\lambda_5 & 0\\
0 & 0 & 0 & \lambda_3-\lambda_5
\end{pmatrix},
\nonumber\\
Z={}&
\begin{pmatrix}
\lambda_1 & \lambda_5 & \sqrt{2}\lambda_6\\
\lambda_5 & \lambda_2 & \sqrt{2}\lambda_7\\
\sqrt{2}\lambda_6 & \sqrt{2}\lambda_7 & \lambda_3+\lambda_4
\end{pmatrix},
 \label{eq:2hdm-pu-matrices}
\end{align}
\endgroup
where $\mathrm{eig}\, A$ denotes the eigenvalues of a matrix $A$.

To derive the tree-level BFB condition, we rewrite the
quartic part of the potential, $V_4$, as
\begin{align}
 \frac{V_4}{(\Phi_1^\dagger\Phi_1)^2}
={}&\frac{\lambda_1}{2}+\frac{\lambda_2}{2}x^4+\lambda_3x^2
+x^2\rho^2(\lambda_4+\lambda_5\cos2\theta)
+2(\lambda_6x+\lambda_7x^3)\rho\cos\theta,
\label{eq:bfb-general}
\end{align}
where $x=(\Phi_2^\dagger\Phi_2/\Phi_1^\dagger\Phi_1)^{1/2}$ and
$z=\rho\cos\theta$ parameterizes the relative direction
$\Phi_1^\dagger\Phi_2=
(\Phi_1^\dagger\Phi_1\Phi_2^\dagger\Phi_2)^{1/2}\rho e^{i\theta}$
with $x\geq0$, $0\leq\rho\leq1$, and $0\leq\theta<2\pi$.
The BFB condition is equivalent to requiring it to be non-negative in all directions spanned by $x,\rho,\theta$.
Thus we impose
\begin{align}
  \min_{x,\rho,\theta} \frac{V_4}{(\Phi_1^\dagger\Phi_1)^2} \geq 0 \, .
\end{align}

\refstepcounter{section}
\label{sec:rge}
\noindent\textit{Appendix~\thesection: Renormalization-group equations.}---We summarize the one-loop RG equations used in the numerical analysis.
We work in the CP-conserving limit and neglect all fermion Yukawa
couplings except for the top-quark Yukawa coupling.
We denote by $Y_t$ the coupling appearing in
$\overline{Q}_L\widetilde{\Phi}_2 t_R$.
The radiatively induced Yukawa couplings to the other Higgs doublet
start at higher loop order and are therefore not included in the
one-loop running considered here.
The equations follow the standard general-2HDM RG equations
in our convention for the potential~\cite{Oredsson:2018yho}.

For convenience, we define
\begin{align}
 {\cal D} \equiv 16\pi^2\frac{d}{d\log\mu},
 \qquad
 G\equiv 3g^2+g'^2 .
 \label{eq:rge-def}
\end{align}
The beta functions for the quartic couplings are
\begin{align}
{\cal D}\lambda_1={}&
 12\lambda_1^2
 +4\lambda_3^2
 +4\lambda_3\lambda_4
 +2\lambda_4^2
 +2\lambda_5^2
 +24\lambda_6^2
\nonumber\\
&+\frac{3}{4}
 \left(3g^4+g'^4+2g^2g'^2\right)
 -3G\lambda_1 ,
\label{eq:rge-lambda1}
\\
{\cal D}\lambda_2={}&
 12\lambda_2^2
 +4\lambda_3^2
 +4\lambda_3\lambda_4
 +2\lambda_4^2
 +2\lambda_5^2
 +24\lambda_7^2
\nonumber\\
&+\frac{3}{4}
 \left(3g^4+g'^4+2g^2g'^2\right)
 -3G\lambda_2
 +12Y_t^2\lambda_2
 -12Y_t^4 ,
\label{eq:rge-lambda2}
\\
{\cal D}\lambda_3={}&
 (\lambda_1+\lambda_2)(6\lambda_3+2\lambda_4)
 +4\lambda_3^2
 +2\lambda_4^2
 +2\lambda_5^2
\nonumber\\
&+4(\lambda_6^2+\lambda_7^2)
 +16\lambda_6\lambda_7
 +\frac{3}{4}
 \left(3g^4+g'^4-2g^2g'^2\right)
\nonumber\\
&-3G\lambda_3
 +6Y_t^2\lambda_3 ,
\label{eq:rge-lambda3}
\\
{\cal D}\lambda_4={}&
 2(\lambda_1+\lambda_2)\lambda_4
 +8\lambda_3\lambda_4
 +4\lambda_4^2
 +8\lambda_5^2
\nonumber\\
&+10(\lambda_6^2+\lambda_7^2)
 +4\lambda_6\lambda_7
 +3g^2g'^2
 -3G\lambda_4
 +6Y_t^2\lambda_4 ,
\label{eq:rge-lambda4}
\\
{\cal D}\lambda_5={}&
 (2\lambda_1+2\lambda_2+8\lambda_3+12\lambda_4)\lambda_5
\nonumber\\
&+10(\lambda_6^2+\lambda_7^2)
 +4\lambda_6\lambda_7
 -3G\lambda_5
 +6Y_t^2\lambda_5 ,
\label{eq:rge-lambda5}
\\
{\cal D}\lambda_6={}&
 2(6\lambda_1+3\lambda_3+4\lambda_4+5\lambda_5)\lambda_6
\nonumber\\
&+2(3\lambda_3+2\lambda_4+\lambda_5)\lambda_7
 -3(G-Y_t^2)\lambda_6 ,
\label{eq:rge-lambda6}
\\
{\cal D}\lambda_7={}&
 2(6\lambda_2+3\lambda_3+4\lambda_4+5\lambda_5)\lambda_7
\nonumber\\
&+2(3\lambda_3+2\lambda_4+\lambda_5)\lambda_6
 -3(G-3Y_t^2)\lambda_7 .
\label{eq:rge-lambda7}
\end{align}

The quadratic parameters obey
\begin{align}
{\cal D}m_{11}^2={}&
 6\lambda_1m_{11}^2
 +(4\lambda_3+2\lambda_4)m_{22}^2
 -12\lambda_6m_{12}^2
 -\frac{3}{2}Gm_{11}^2 ,
\label{eq:rge-m11}
\\
{\cal D}m_{22}^2={}&
 (4\lambda_3+2\lambda_4)m_{11}^2
 +6\lambda_2m_{22}^2
 -12\lambda_7m_{12}^2
\nonumber\\
&-\frac{3}{2}Gm_{22}^2
 +6Y_t^2m_{22}^2 ,
\label{eq:rge-m22}
\\
{\cal D}m_{12}^2={}&
 -6(\lambda_6m_{11}^2+\lambda_7m_{22}^2)
\nonumber\\
&+\left(
 2\lambda_3+4\lambda_4+6\lambda_5
 -\frac{3}{2}G+3Y_t^2
 \right)m_{12}^2 .
\label{eq:rge-m12}
\end{align}

For completeness, the gauge couplings and the top Yukawa coupling
are evolved according to
\begin{align}
 {\cal D}g' = 7g'^3 ,\quad
 {\cal D}g= -3g^3 , \quad
 {\cal D}g_s = -7g_s^3 ,
\label{eq:rge-gauge}
\end{align}
\begin{align}
 {\cal D}Y_t &=
 Y_t\left(
 \frac{9}{2}Y_t^2
 -8g_s^2
 -\frac{9}{4}g^2
 -\frac{17}{12}g'^2
 \right).
\label{eq:rge-yt}
\end{align}

The origin of the radiative violation of the FSR is particularly
transparent from Eqs.~\eqref{eq:rge-lambda6} and
\eqref{eq:rge-m12}.  At the matching scale, where
$\lambda_6(\Lambda)=m_{12}^2(\Lambda)=0$, one finds
\begin{align}
 \left.{\cal D}\lambda_6\right|_{\mu=\Lambda}
 &=
 2(3\lambda_3+2\lambda_4+\lambda_5)\lambda_7 ,
\label{eq:rge-source-lambda6}
\\
 \left.{\cal D}m_{12}^2\right|_{\mu=\Lambda}
 &=
 -6\lambda_7m_{22}^2 .
\label{eq:rge-source-m12}
\end{align}
Thus, a nonzero $\lambda_7$ acts as a source for both
$\lambda_6$ and $m_{12}^2$ below the matching scale, even though
they vanish at $\mu=\Lambda$.

\end{document}